\documentclass{article}
\usepackage[utf8]{inputenc}
\usepackage{csquotes}
\usepackage[english]{babel}

\title{Boosting quantum key distribution via the end-to-end physical control}
\author{
    A.\,D.\,Kodukhov$^{1,2}$,
    V.\,A.\,Pastushenko$^{1,2}$,
    N.\,S.\,Kirsanov$^{1,2,3}$,
    D.\,A.\,Kronberg$^{1,2,4}$,\\  
    V.\,M.\,Vinokur$^{1,5}$,
    M.\,Pflitsch$^{1}$,
    G.\,B.\,Lesovik$^{1,2}$
   }
\date{\normalsize{
    $^1$Terra Quantum AG, St. Gallerstrasse 16A, CH-9400 Rorschach, Switzerland\\
    $^2$Moscow Institute of Physics and Technology, 141700, Institutskii Per. 9, Dolgoprudny, Moscow Distr., Russian Federation;\\
    $^3$Low Temperature Laboratory, Department of Applied Physics, Aalto University, Espoo, Finland;\\
    $^4$Department of Mathematical Methods for Quantum Technologies, Steklov Mathematical Institute of Russian Academy of Sciences, Gubkina str. 8, Moscow 119991, Russia;\\
    $^5$Physics Department,
    City College of the City University of New York,
    160 Convent Ave, New York, NY 10031, USA 
    \\
    }
    \today
}

\usepackage{authblk}
\usepackage{indentfirst} 
\usepackage[unicode, pdftex]{hyperref}
\usepackage{braket}
\usepackage{graphicx}
\DeclareGraphicsExtensions{.pdf,.png,.jpg}
\usepackage[10pt]{extsizes}
\usepackage[left=22mm, top=16mm, right=22mm, bottom=20mm, nohead, footskip=10mm]{geometry}
\usepackage{amssymb}
\usepackage{amsmath}
\DeclareUnicodeCharacter{0301}{\'{e}}
\usepackage[usenames]{color}
\usepackage[dvipsnames]{xcolor}
\usepackage{colortbl}

\usepackage[
    backend=bibtex, 
    natbib=true,
    style=phys,
    biblabel=brackets,
    giveninits=true,
    abbreviate=false,
    doi=false, 
    url=true,
    isbn=false,
    block=space,
    backref=false, 
]{biblatex}
\DeclareFieldFormat
  [article,inbook,incollection,inproceedings,patent,thesis,unpublished]
  {title}{#1\isdot}
\DeclareFieldFormat[article,inproceedings,incollection]{journaltitle}{\textit{#1}}
\begin{document}
\maketitle

\begin{abstract}
Quantum key distribution (QKD) is a cornerstone of the secure quantum encryption. 
Building on the quantum irreversibility, we develop a technique reborning the existing QKDs into protocols that are unrestricted in distance and have unprecedented high rates enhanced up to the standard protocols communication speeds.
The core of our method is the continuous end-to-end physical control of information leaks in the quantum channel. Contrary to the existing long-distance QKD offerings, our technique does not require any trust nodes.
 
\end{abstract}
\bigskip
\section{Introduction}
Quantum cryptography is the response to the quest of securing the storage and transfer of information in the face of the upcoming quantum future. 
A keystone of quantum cryptography is quantum key distribution (QKD) offering the capability for symmetric generation of secret bit sequences under the protection of limited distinguishability of non-orthogonal quantum states, no-cloning property of quantum systems, and quantum entanglement.
Adding to the protecting protocol the techniques based on the one-time-pad encryption which are impossible to break even with a quantum computer makes novel QKD algorithms a rescue against the looming quantum computers threat and a roadway to tomorrow's unconditionally secure cryptography.
 
Yet the existing QKD schemes are still far from being a complete miraculous remedy as their key distribution rates and achievable distances do not meet all the practical needs. There is fundamental bound \cite{TGW, PLOB} for conventional quantum cryptography performance in lossy communication lines which makes it impractical for long distances. Possible solutions are whether not scalable \cite{TwinField} or need quantum repeaters which are yet relatively difficult to use in practice. Another possible  recipe for enhancing the performance of the practical QKD implementations is the installment of the trusted equipment in-between the legitimate users.
This, however, inevitably compromises security as it creates a back door to information processed in the intermediate trusted nodes.
Here, we introduce a method that enables a drastic improvement of the transmission rates without compromising the transmission distances of any existing QKD protocols based on quantum irreversibility without the necessity of introducing the trusted in-line devices.

Common surmises making a base of the standard quantum cryptography can be formulated as follows:
\begin{displayquote}
\begin{enumerate}
    \item \textit{All the quantum control over the transmitted information is executed by the recipient (Bob) measuring the received quantum state. Importantly, the physical control over the line is not included into the control procedure.}
    \item \textit{All the losses in the quantum channel can be collected by an eavesdropper (Eve) enabling her to access the informational content of the leaked fraction of the signal.}
\end{enumerate}    
\end{displayquote}
However, in reality, most of the losses of the transmitted signal are related to the signal's scattering on the imperfections distributed along the transmission channel. 
This process is physically irreversible, the quantum mechanism of the irreversibility being revealed in\,\cite{H_teorem,arrow}.
This irreversibility does not leave Eve with much freedom in processing scattered signal wave packets. 
Eve may directly collect and measure the dispersed components of a scattered wave packet but with each component on its own carrying scarce statistics, her precision would be completely obscured by the quantum noise.
Eve thus would tend to unify losses into the, desirably, one narrow wave packet by reversing the scattered wavefront. 
However, no universal quantum operation allowing to reverse an arbitrary wave function does exist.
As was shown in Refs.\,\cite{H_teorem,arrow,entropy_dynamics,H_theorem_demon}, even to construct an operation reversing a simplest specific scattering event and, consequently, to harvest information, Eve already has to have an\,\textit{a priori} information about the scattered signal.
And if a single scattering occurs once per a wavelength, then to handle scattering in a 1\,km-long channel, Eve would need to build a billion quantum Maxwell demon-like devices, each reversing the entropy dynamics of losses from the corresponding individual scattering center.

All the above implies that extracting information from the losses due to scattering, to which we can refer as to dissipative natural losses, is unfeasible. Let alone that even collecting the scattered radiation requires an implausible antenna covering a significant section of the transmission channel. 
This, in turn, means that the factual premise for quantum cryptography is 
\begin{displayquote}
    \textit{Eve can only obtain information from the reversible losses which usually have a local character.}
\end{displayquote}
Therefore, to ensure eavesdropping, Eve has to arrange deliberate leakages in the channel, as
we have already discussed in\,\cite{long-distance}.

\begin{figure}[h]
    \centering
    \includegraphics[scale=0.41]{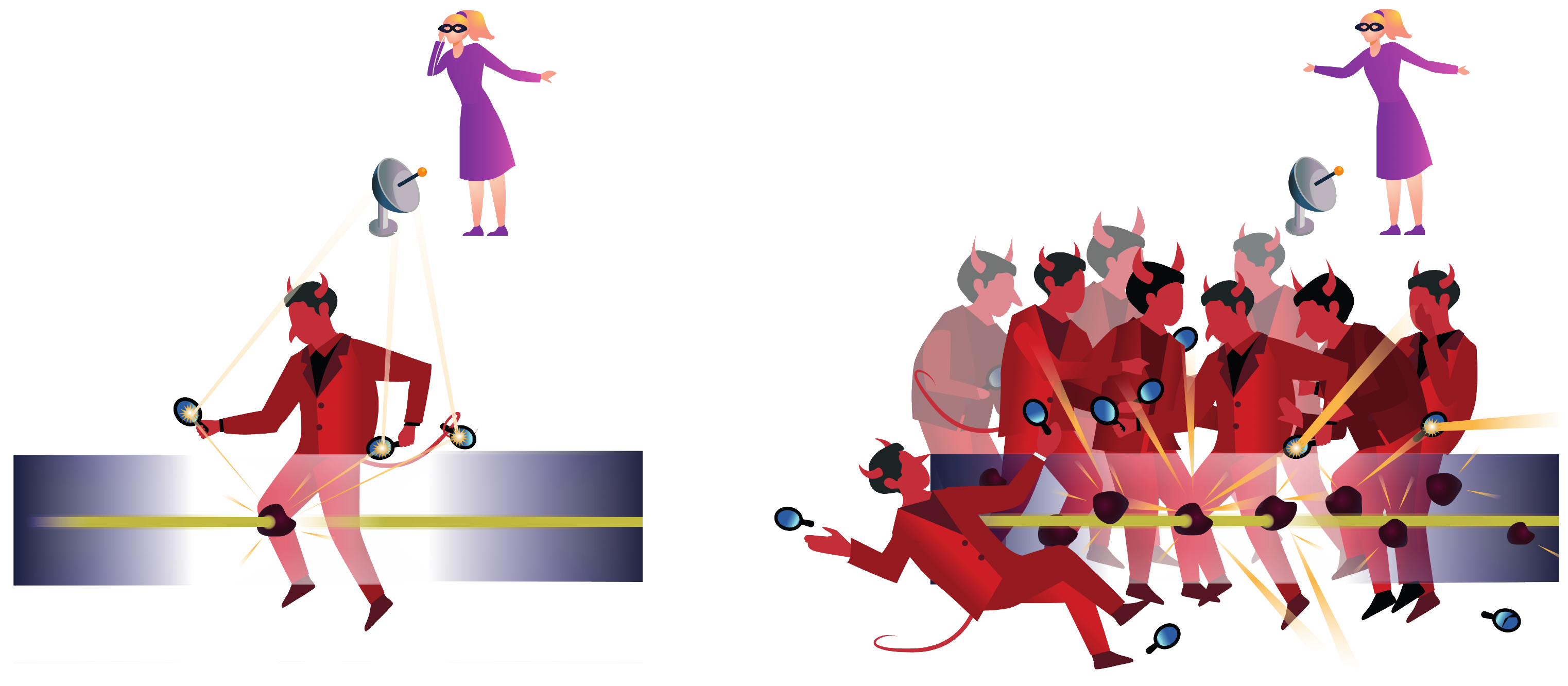}
    \caption{It takes a Maxwell demon-like device to get information from the radiation scattered on a local homogeneity of optical fiber. For 1 km-long fiber, the required number of demons reaches $10^9$, thus, making collecting information from scattering infeasible.}
    \label{key rate}
\end{figure}

The protection method that we propose is based on the permanent control and analysis of losses in the quantum channel with the response adapting the QKD setup parameters. 
We demonstrate how using our method one can significantly enhance the efficiency of the QKD protocols.
In particular, we examine the enhancement of the Decoy-State BB-84\,\cite{Decoy1,Decoy2} and Coherent One Way (COW)\,\cite{OriginalCOW} protocols and show numerically  the remarkable gains in their efficiency in terms of the key generation rate and the achievable distance between the legitimate users. 
In the context of the line control technique, the differential phase shift (DPS) QKD \cite{DPS,DPS2} appeared to be identical to COW.
Moreover, we extend our findings onto other protocols and demonstrate that in certain cases our modifications can substantially reduce the complexity of the necessary equipment.

\section{Method description}

We develop our technique in the context of the prepare-and-measure QKD protocols, although it applies to other protocols as well.
Typically, Alice encodes a random bit sequence into the quantum states constituting the signal and transmits them via a quantum channel to Bob, who acquires this sequence (the raw key) using quantum measurements. The raw key is then processed to eliminate all errors and information that leaked to Eve.
Our technique enables legitimate users to control the unauthorized connections to the quantum channel and to correspondingly adapt a QKD protocol to multiply its effectiveness. 
This adaptation particularly addresses the stage of the post-processing of the raw key.
Although our technique also implies adapting the parameters of the quantum signals themselves, even its implementation at the post-processing level solely drastically boosts the QKD characteristics.

Our technique is based on the constant control of the quantum channel's losses using the reflectometry, i.e., analysis of the reflected test signals, and the transmittometry, the analysis of the transmitted test signals, the methods, which we describe in detail below.
This allows for Alice and Bob to precisely learn what proportion of the signal has potentially leaked to Eve and to distinguish the leaks from the dissipative scattering which cannot be deciphered.
Successful realization of the line control requires the evaluation of the initial loss profile of the quantum channel based on the condition that at the preliminary stage the signal is not intercepted.
Knowing the exact proportion of the stolen signal, Alice and Bob can accurately estimate their informational advantage over Eve and choose the best signal characteristics to further maximize this advantage.
An accurate analysis of the information advantage brings the significant enhancement of the key distribution rate as it allows for performing better error correction and privacy amplification procedures. 
The latter serves to effectively diminish the information available to Eve without sacrificing much too many bits of the raw key.

\section{Line control}

As mentioned above, the Alice/Bob party can distinguish whether the losses occurring in the optical fiber are of a dissipative nature. 
As it has been mentioned in the Introduction, it is technically impossible for an eavesdropper to extract information about the transmitted raw key bits from the dissipative losses. 
This and the subsequent adapting the transmission making deciphering impossible marks a paradigm shift in the approach to the QKD protocol and enables increasing the key generation rate in the framework of the innovative emergent versions of these protocols. Further improvements rest on estimates of the level of the information an eavesdropper can obtain from non-dissipative local losses associated with the specifics of routing (they are usually referred to as events). Those include losses on connectors, welds, bends and cracks.
They may occur both naturally during, e.g., the optical fiber installation as well as result from the eavesdropper's activity.

\subsection{Reflectometry}
\begin{figure}[h]
\begin{minipage}[h]{0.49\linewidth}
    \center{\includegraphics
    [width=0.97\linewidth]{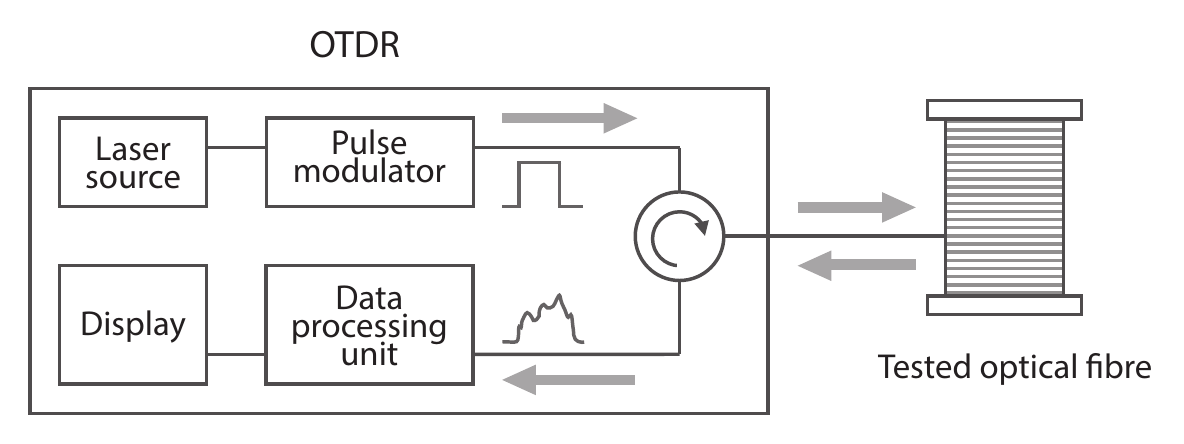}
    \caption{A sketch of an optical time-domain reflectometer (OTDR). The QTDR consists of the laser generating the input signal, the examining optical fiber modifying the input pulse into the back-scattered signal, and of the data-analyzing unit, producing a reflectogram.}
    \label{reflectometer_scheme}}
\end{minipage}
\hfill
\begin{minipage}[h]{0.49\linewidth}
    \center{\includegraphics
    [width=0.97\linewidth]{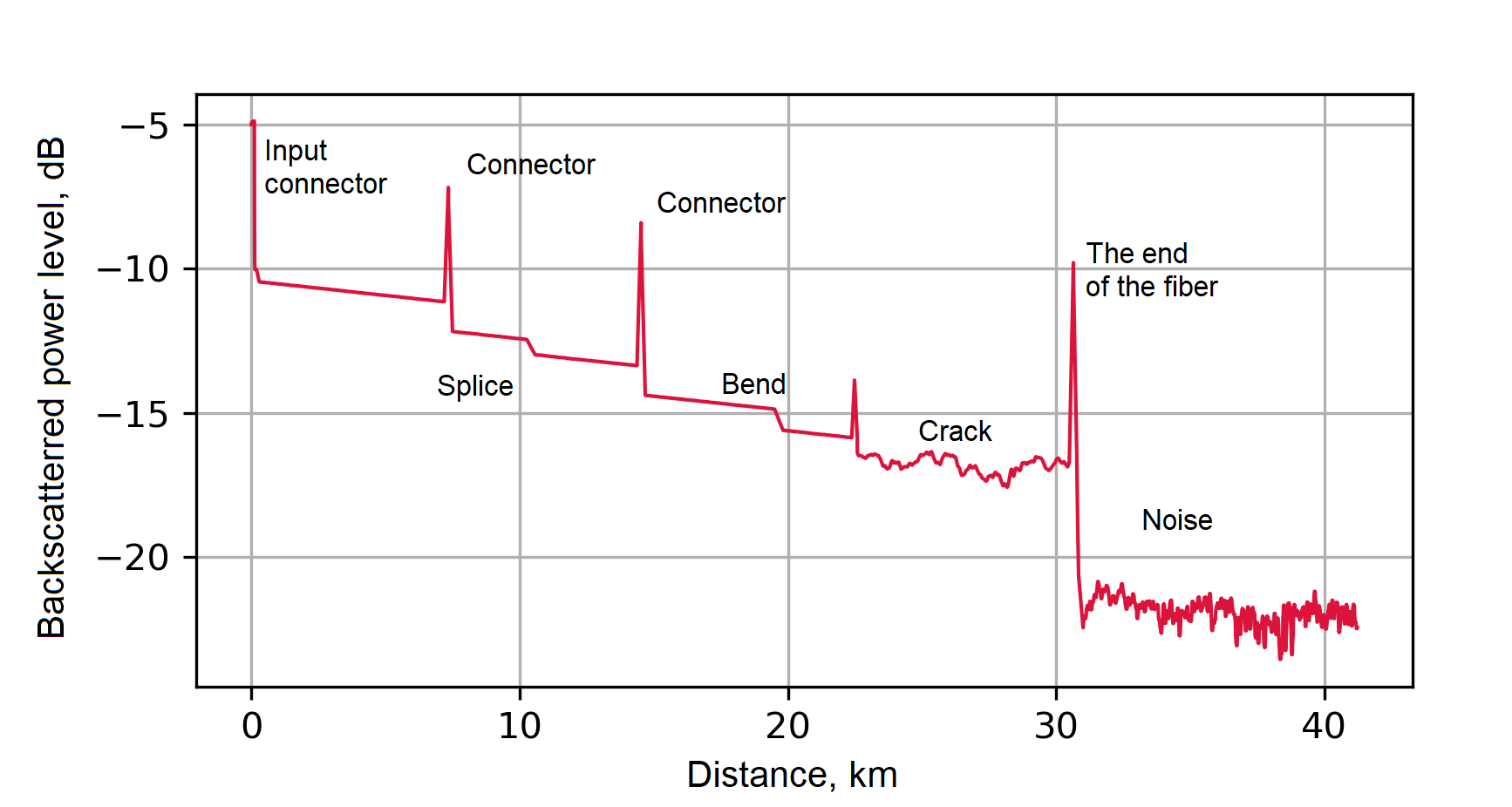} 
    \caption{An example of a reflectogram obtained by detecting and analyzing the back-scattered signal. The reflectograms display the losses associated with every event in the fiber, i.e., losses on connectors, bends, cracks, and splicing.}
    \label{example_reflectograms}}
\end{minipage}
\end{figure}

The main contribution to the intrinsic losses in an optical fiber line comes from the Rayleigh scattering, caused, for example, by irregularity in the optical fiber density and does not exceed 0.2 dB/km for modern fibers, and from the specifics of routing, i.e., losses on connectors, welds, bends, and cracks.
These losses (events) can be detected using optical reflectometers. 
The optical reflectometry is based on the registration of the back-scattered optical radiation resulting from the test impulses propagating through the fiber, see Fig.\,\ref{reflectometer_scheme}. 
Measuring the delays of the signals' arrival, one can calculate the distances to the spots of the local losses occurrences. Moreover, specific features of the back-scattered signal allow for determining the type of each event. 
Small imperfections that cause Rayleigh scattering are homogeneously spread along the fiber and correspond to the linear parts of the potential reflectogram, see Fig.\,\ref{example_reflectograms}. 
The slope of these parts is equal to the fiber's losses parameter\,dB/km. 
The essential local defects cause a sharp local decline of the intensity and comply with other sections of the reflectogram including narrow peaks and sharp linear drops. 

To distinguish the intrinsic losses from those that appear due to the eavesdropper's activity, the naturally occurring events are pre-detected with the help of the optical reflectometer and documented.
Since, as we have already shown, it is not feasible for an eavesdropper to have the dissipative losses effectively exploited  for extracting information about the raw key, an eavesdropper is left with the necessity of arranging the attacks provoking the non-dissipative local losses.
The most general eavesdropping attack implies conducting a general unitary transformation on the propagating signal's system combined with an ancillary one.
However, the only way to couple the systems is to introduce significant alterations to the optical fiber medium which would inevitably cause new events on the reflectogram and would be detected.

Any intended eavesdropping attack would the creation of the leakage point in the fiber channel. 
A proper installation of an optical fiber guarantees the absence of defects producing notable spikes in a reflectogram.
Thus, any new defect of the fiber channel that diverts an appreciable part of the propagating signal leads to a  noticeable distortion of the reflectogram.
Accordingly, upon discovering such a distortion, legitimate users would terminate the protocol, depriving an eavesdropper of the possibility of conducting a variety of hacking attacks like, for instance, the intercept-resend action. 

\subsection{Transmittometry and the accuracy limit of the line control}
Another approach to exercising the line control and detecting the eavesdropper's activity is the transmittometry. 
The method utilizes the fact that besides the signal pulses that carry the information about the raw key, Alice has to send special test pulses. 
Test pulses do not carry any information about the sharing secret bit string, but they are supposed to have the highest possible intensity which provides the highest precision of the leakage detection. 
The parameters of the prepared test pulse, such as intensity, phase, length, and shape of the pulse, are chosen randomly by the sender. 
Alice sets the parameters according to an auxiliary pre-generated random bit string that remains secret and unknown even to Bob. 
Upon the transmission of the test pulse, and after Bob has measured it, Alice announces the chosen settings. Then the legitimate users, like Bob, implement the cross-check of parameters and determine the losses in the channel. 
The natural losses should be premeasured and also documented to allow for detecting eavesdropper's activity. 
Importantly, the time intervals for the test pulses transmission are chosen according to some short preshared secret key, which part is also utilized for the authentication, and, thus, are unknown to Eve. 
This guarantees that Eve, who does not know in advance whether a pulse is a signal or a test one, does not have an opportunity of deliberate treating these two types of pulses differently.
This procedure combined with the limitations that reflectometry methods impose on Eve's actions makes it impossible for her to create an additional permanent leakage constant or a leakage targeting some chosen signal and test pulses.
Thus, legitimate users can evaluate Eve's knowledge about signal pulses exploiting the information that they get analyzing the test pulses.

To evaluate the artificially created leakage $r_E$, the legitimate users have to analyze the non-scattered part of intensive test signals that propagate along the whole optical channel from Alice to Bob. 
Since the leakage is constant, the evaluation of $r_E$ obtained through test pulses is applicable to information-carrying signals. 
The maximal achievable precision of detecting $r_E$ is naturally bounded by the magnitude of the Poisson noise. 
If Alice sends a test pulse which is a coherent state with an average number of photons equal to $n^A_{\text{test}}$, then Bob receives a coherent state with $n^B_{\text{test}}=T\cdot n^A_{\text{test}}$. 
Where $T=10^{-\mu D}$ is the transmittance of the  whole line of the length $D$ with the parameter of losses equal to $\mu$ (typically, in optical fibers $\mu=1/50\mbox{ km}^{-1}$).
For the fluctuations of the coherent states 
\begin{gather}
    \delta n_{\text{test}}^B
    \sim
    \sqrt{n_{\text{test}}^B}
    \sim
    \sqrt{T \cdot n_{\text{test}}^A}.
\end{gather}
Thus, the minimum detectable $r_E$,
\begin{equation}
    r_{E,\min}
    \sim
    \frac{\delta n^B_{\text{test}}}{n^B_{\text{test}}}
    \sim
    \frac{1}{\sqrt{T\cdot n^A_{\text{test}}}}
    =
    \sqrt{\frac{10^{\,0.02 D}}{n^A_{\text{test}}}}\,.
\end{equation}
To reach such an accuracy of the line control, the average photon number in the test pulses should be
\begin{equation}
    n^A_{\text{test}}
    \sim
    \frac{10^{\,0.02 D}}{r^2_{E,\min}}\,.
    \label{re}
\end{equation}

\begin{figure}[h]
    \centering
    \includegraphics[scale=0.43]{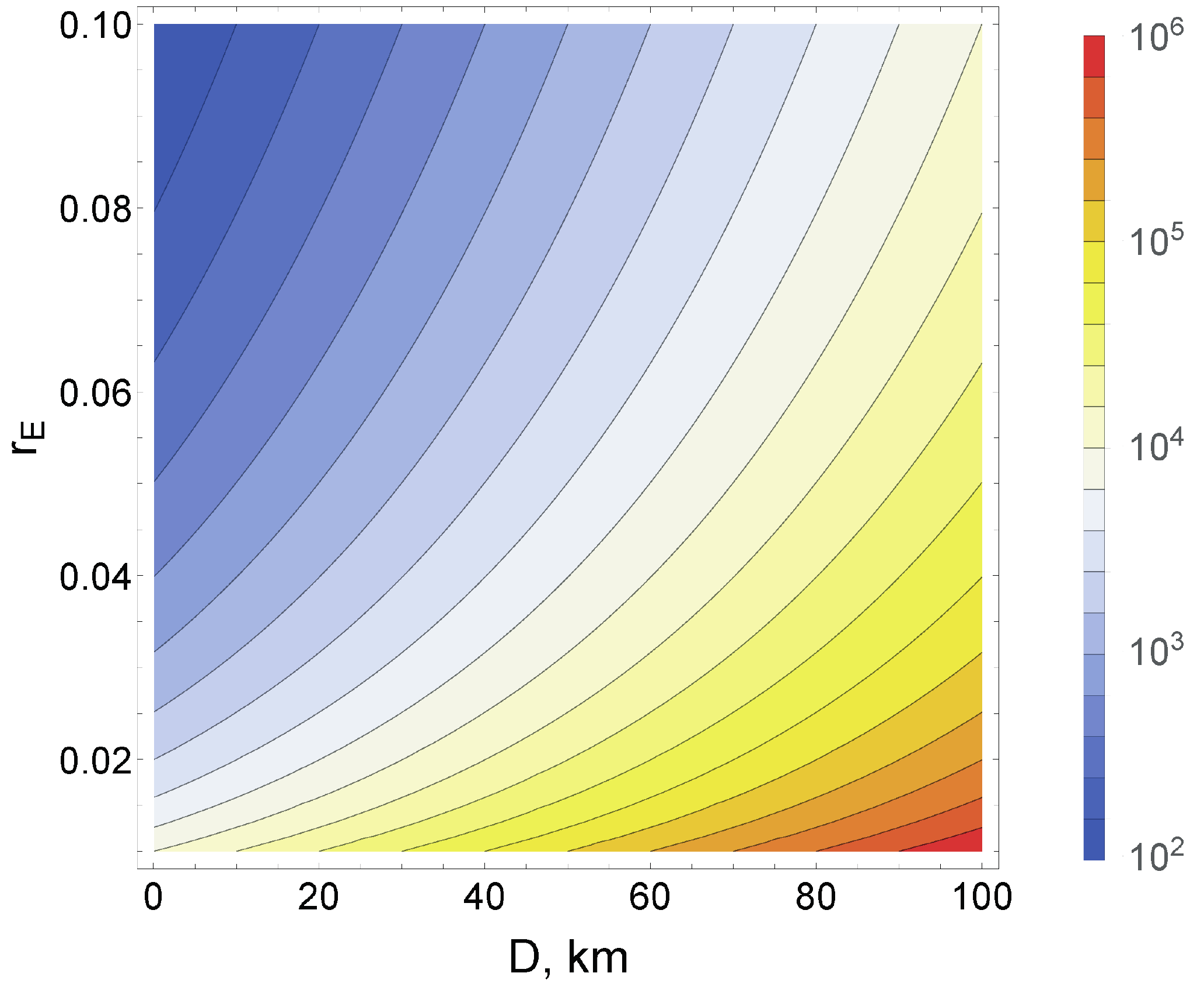}
    \caption{The intensity of the test pulses $\langle n_{\text{test}}^A\rangle$ as a function of $r_E$ and $D$.}
    \label{test_pulses}
\end{figure}

Therefore, the desired precision of the leakage detection establishes a lower bound of the test pulses' intensity that should be exploited.
Fig.\,\ref{test_pulses} represents the average photon number in the test pulses as a function of the distance between the legitimate users $D$ and the minimum detectable leakage $r_E$. 
To achieve the precision of the leakage about $1\%$ at the distance of 100\,km one has to utilize the test pulses with no less than $10^6$ photons, which can be easily implemented experimentally.  

\newpage
\section{Improvement of QKD protocols}
Depending on whether or not a QKD protocol includes the phase randomization of the employed quantum states, one can classify them into two groups: protocols without phase randomization utilizing pure coherent states, and protocols with applicable single-photon security proofs.

In protocols without phase randomization, the coding is described in the phase space one-mode of coherent states. 
On occasions, the second mode is utilized as a reference to simplify measurements.
Exemplary representatives of this sort of protocols are Coherent One-Way\cite{OriginalCOW,COW_record,COW2009}, Differential Phase Shift \cite{DPS,DPS2,DPS_record}, and Y-00 \cite{Y-00,noise_prtotected,barbosa}. 
All the listed protocols have already been successfully realized experimentally.

The second group includes protocols with phase randomization and exploits the qubit (or qudit) space to describe the coding configuration.
Although ideally, the protocols are to be realized via the single-photon states, they are practically implemented as multi-photon pulses, for example, the phase-randomized weak coherent states.
A typical example of such a protocol is the BB-84 improved with the Decoy-State method\,\cite{Decoy1,Decoy2,bb84_decoy_experimental}. 
The ``4+2" QKD protocol\,\cite{4+2}, the Six-State protocol\,\cite{six-state}, T-12\,\cite{T12,T12_experiment} and the SARG04 QKD protocol \cite{SARG04,sarg} can also be classified under the group.

Remarkably, the proposed technique can be used not only for securing communications via the fiber optic channels but also -- with the proper modifications -- to protect the communications across the free space, like, for example, the ground-satellite communications.
In the case of the satellite-based QKD schemes\,\cite{Satellite_1,Satellite_2,Satellite_3}, the physical line control would primarily rely on the analysis of the transmitted test signals. 
However, as the device-independent cryptography is still in its infancy, at the moment, any satellite-based scheme implies that the satellite itself is a trusted node.
Thus, unlike fiber-based communications, the use of satellites cannot yet guarantee full communication security.

\subsection{BB84 protocol}
In this section, we address an exemplary protocol exercising phase randomization, the Decoy-State BB84 protocol, and demonstrate the gain in the key generation rate which can be provided by the line control.
To show the benefits brought by the line control, we must calculate the key generation rates both for the original Decoy-State BB84 and its line control modification.
In the case of the latter, this task is rather straightforward, as for it we must consider only the one relevant type of attack exploiting the non-dissipative local losses in the quantum channel.
In contrast, the accurate rate evaluation for the original protocol requires taking into account a spectrum of various attacks but this is not needed for our illustrative purposes -- we will only estimate the upper bound for the original protocol's rate.
The estimate in such a case is yielded from the analysis of the standard photon number splitting (PNS) attack with the eavesdropper getting all signal lost in the channel.
A conventional approach to Quantum Cryptography implies that the attack is possible.

In the case of the PNS attack on the original protocol, due to substantial channel losses, Eve knows large amount of information about the raw key right before the privacy amplification stage.
This means that the privacy amplification should largely reduce the key length meaning relatively low value of the upper bound of the key generation rate for the BB84 protocol.
In the case of the modified protocol, Eve gets just a small fraction of propagating signals and, thus, obtains less information about the raw key.
This provides legitimate users with a much higher key rate because the privacy amplification is not as severe as in the original Decoy-State BB84.

Note that in the original Decoy-State BB84, only single-photon pulses are involved in the key generation process, while all multi-photon pulses considered as insecure. 
Additional decoy states are used to estimate the fraction of the raw key bits obtained from true single-photon pulses.
The signal's intensity should be close to 1 photon per pulse to increase the percentage of single-photon states.
Fortunately, line control allows for using higher intensities resulting in a greater key generation rate without losing security.
Moreover, legitimate users can fully replace the Decoy-State method with the line control.

\subsubsection{Decoy-State BB84}

\begin{figure}[h]
    \centering
    \includegraphics[scale=0.8]{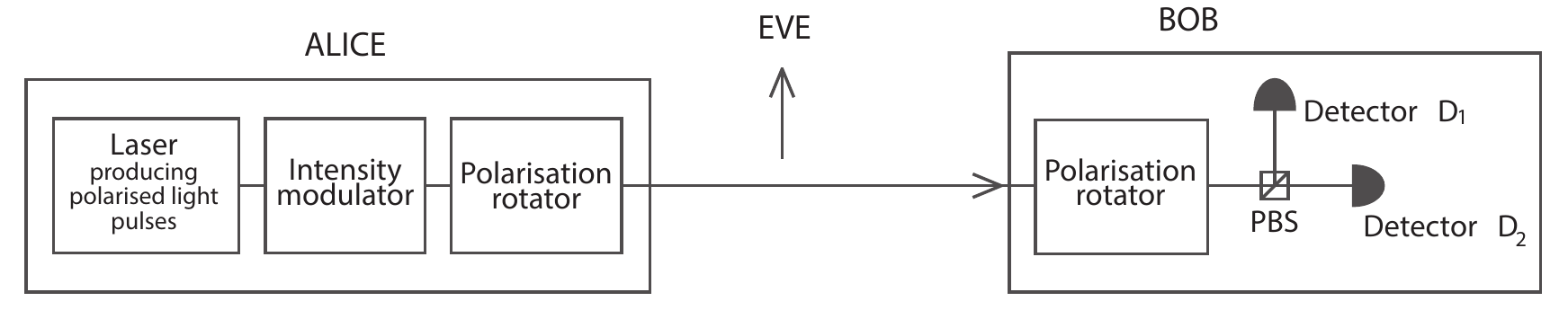}
    \caption{The scheme of Decoy-State BB84 protocol. 
    Alice prepares a signal or a decoy state by utilizing the intensity modulator and uses the polarization rotator to encode the information into the signal states.
    Bob switches between the bases by applying the polarization rotator and performs the measurement by using the polarizing beam-splitter PBS and single-photon detectors $D_1,D_2$. 
    Eve conducts PNS attack.}
    \label{bb84_scheme}
\end{figure}

In the BB84 protocol, Alice encodes every bit of the randomly generated string into one of four quantum states, $\left\{\ket{0_x},\ket{1_x};\ket{0_z},\ket{1_z}\right\}$, forming two mutually unbiased orthonormal bases X and Z. 
The bit ``0"(``1") can be encoded into the $\ket{0_x}(\ket{1_x})$ or $\ket{0_z}(\ket{1_z})$ states, the choice is to be done randomly. 
Bob guesses the basis for each bit with the 1/2 probability for the success and measures the received state in the chosen basis. 
Then, Alice reveals the bases and the legitimate users discard bits with no basis coincidence to obtain a shared raw key.

Originally, the protocol was designed for the single-photon pulse implementation\,\cite{BB_84}. 
In the experimental BB84 realizations, see Fig.\,\ref{bb84_scheme}, an attenuated coherent laser light is utilized as a source of single-photon states. It generates weak coherent pulses of the low intensity $|\gamma|^2$ with the unknown random general phase, which are a statistical mixture of the photon-number sates $\ket{n}$ (Fock states) with the Poisson distribution $P_n=\exp(-|\gamma|^2)\cdot|\gamma|^{2n}/n!$
\begin{gather}
\hat{\rho}=\sum\limits_{n=0}^{\infty}P_n\ket{n}\bra{n}\,.
\end{gather}
As a result, the laser sometimes generates multiphoton pulses, which gives a principal advantage for a potential eavesdropper who can conduct photon number splitting (PNS) attack, obtain all the ``additional" photons and store them in the quantum memory until the bases reconciliation. 
The orthogonality condition $\braket{0_x|1_x}=\braket{0_z|1_z}=0$ gives her a possibility to distinguish between the logical bits ``0" and ``1" without any mistake. 
Thus, only single-photon pulses emitted by Alice's laser guarantee the secure QKD.
The length of the secret key that can be achieved in this case is
\begin{gather}
    L_f=L\cdot\frac{1}{2}\left[Q_1(1 - h_2(e_1))-Q \cdot f(E) \cdot h_2(E)\right],
    \label{DecoyKey}
\end{gather}
where $h_2(x) = -x\log_2 x - (1-x)\log_2 (1-x)$ is the binary entropy, $Q$ is the gain of signal states (the probability that a signal state will be detected by Bob) and $E$ is the QBER (Quantum Bit Error Rate), both of the characteristics can be easily obtained from the experiment; $f(E)\in[0,1]$ is the efficiency of the error-correction procedure. The quantity $Q_1$ is the gain of single-photon states (a joint probability that a single-photon pulse was emitted by Alice and was detected by Bob), $e_1$ is the error rate for single-photon pulses. 
Bob cannot distinguish between photons that originated from the single-photon and multi-photon pulses. 
Thus, the legitimate users cannot obtain $Q_1$ and $e_1$ directly and they have to estimate the quantities. So far, the most efficient method was based on the Decoy-State idea \cite{Decoy1,Decoy2}.
To find an upper bound on (\ref{DecoyKey}), one can use non-negativity of binary entropy $h_2$ and get $L_f\leq L\cdot\frac{1}{2}Q_1$.
Eve's activity causes the decrease of the gain of single-photon states $Q_1$, so it is maximum in the Eve's absence: $Q_1\leq\Tilde{Q}_1=T\cdot |\gamma|^2\exp(-|\gamma|^2)$. 
Consequently, the upper bound for the length of a shared secret is
\begin{gather}
    L_f\leq L\cdot\frac{1}{2}T\cdot |\gamma|^2\exp(-|\gamma|^2)\equiv \Tilde{L}_f.
    \label{BB84 id Rkey}
\end{gather}
One can simply show analytically that the upper estimation (\ref{BB84 id Rkey}) for the original Decoy-State BB84 key rate is maximum for $|\gamma|^2=1$.

\subsubsection{BB84 improved with the line control}

Next, we consider the line control technique applied to the BB84. 
In order to extract any information from the channel, Eve has to inflict local non-dissipative artificial losses.
Let us directly calculate the key generation rate, taking into account that the privacy amplification is not as severe as in the original Decoy-State BB84.
Then, the probability for Bob to obtain a conclusive result is
\begin{gather}
    p'\left(\checkmark \right)=\frac{1}{2}\left[1-\exp\left(-T \cdot \left(1-r_E\right)|\gamma|^2\right)\right],
\end{gather}
where the factor $1/2$ appears because of bases reconciliation. Let us also consider a powerful eavesdropper who has a quantum memory. This means that an eavesdropper may store intercepted photons until bases reconciliation and apply optimal measurement obtaining full information about a bit. Thus, whenever Eve intercepts at least one photon, she knows a bit
\begin{gather}
    \max I'(A,E)=0\cdot P_E(0)+1\cdot P_E(\geq 1),
\end{gather}
$P_E(0)$ and $P_E(\geq 1)=1-P_E(0)$ denote, respectively, the probability that Eve intercepts a vacuum state and the probability that she intercepts any positive number of photons. Due to Poisson statistics $P_E(0)=\exp(-r_E|\gamma|^2)$. Thus, after the post-selection procedure and privacy amplification, the shared bit string has the size
\begin{gather}
    L'_f=p'\left( \checkmark \right)L\cdot\left[1-\max I'(A,E)\right]
    = \frac{1}{2} \cdot L \left[ 1-\exp\left(-T \cdot \left(1-r_E\right) \cdot |\gamma|^2\right) \right] \cdot \exp(-r_E \cdot |\gamma|^2).
    \label{BB84lc L}
\end{gather}

\begin{figure}[h]
\begin{minipage}[h]{0.47\linewidth}
    {\includegraphics
    [width=0.9\linewidth]{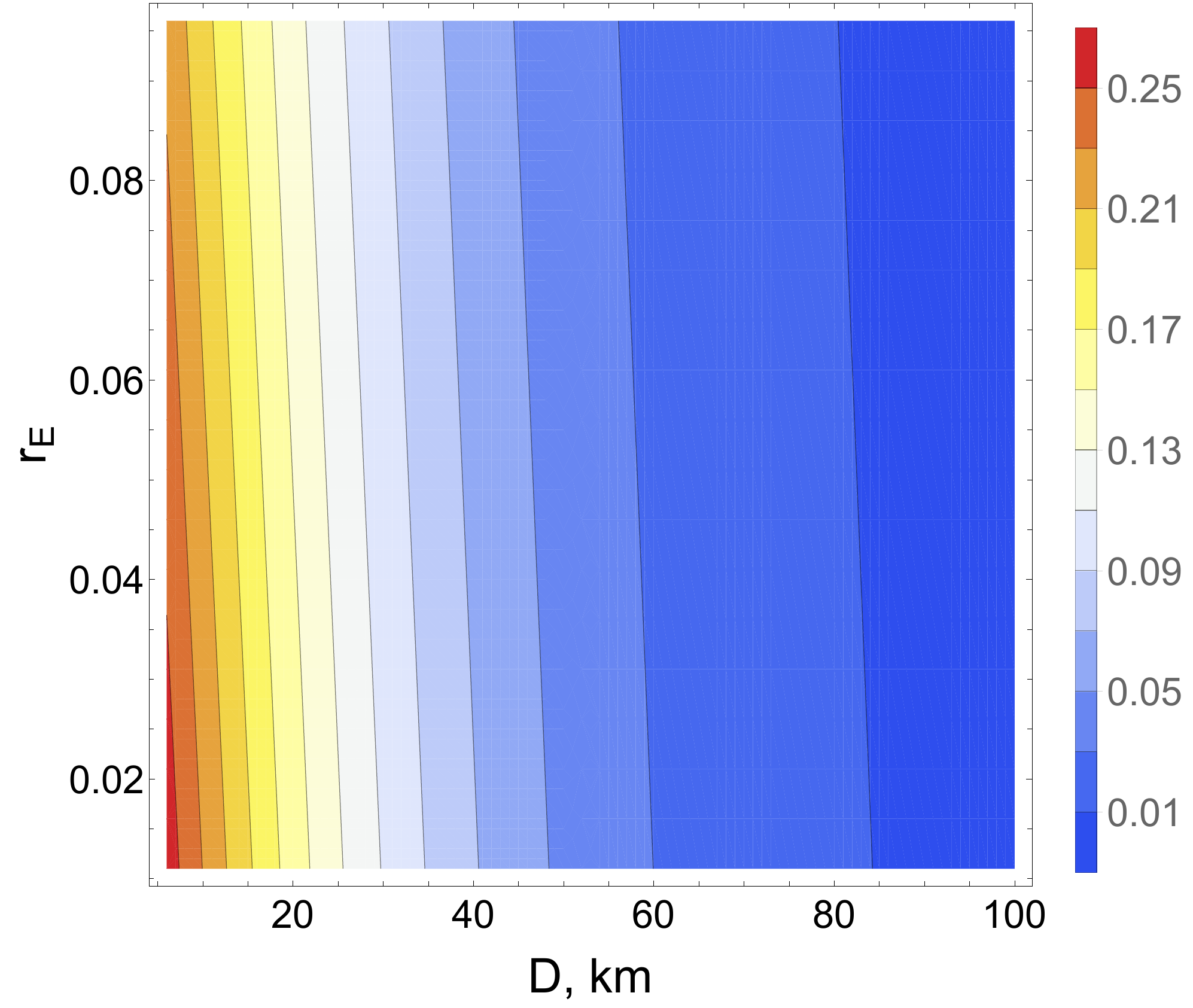}
    \caption{The color plot for the key rate (\ref{BB84lc L}) of the BB84 protocol with the line control as a function of $D$, the distance between Alice and Bob, and $r_E$ if the protocols parameters (namely, signal pulses' intensities) are preserved.}
    \label{bb84-key-rate-no-intensity-optimisation}}
\end{minipage}
\hfill
\begin{minipage}[h]{0.47\linewidth}
    \center{\includegraphics
    [width=0.9\linewidth]{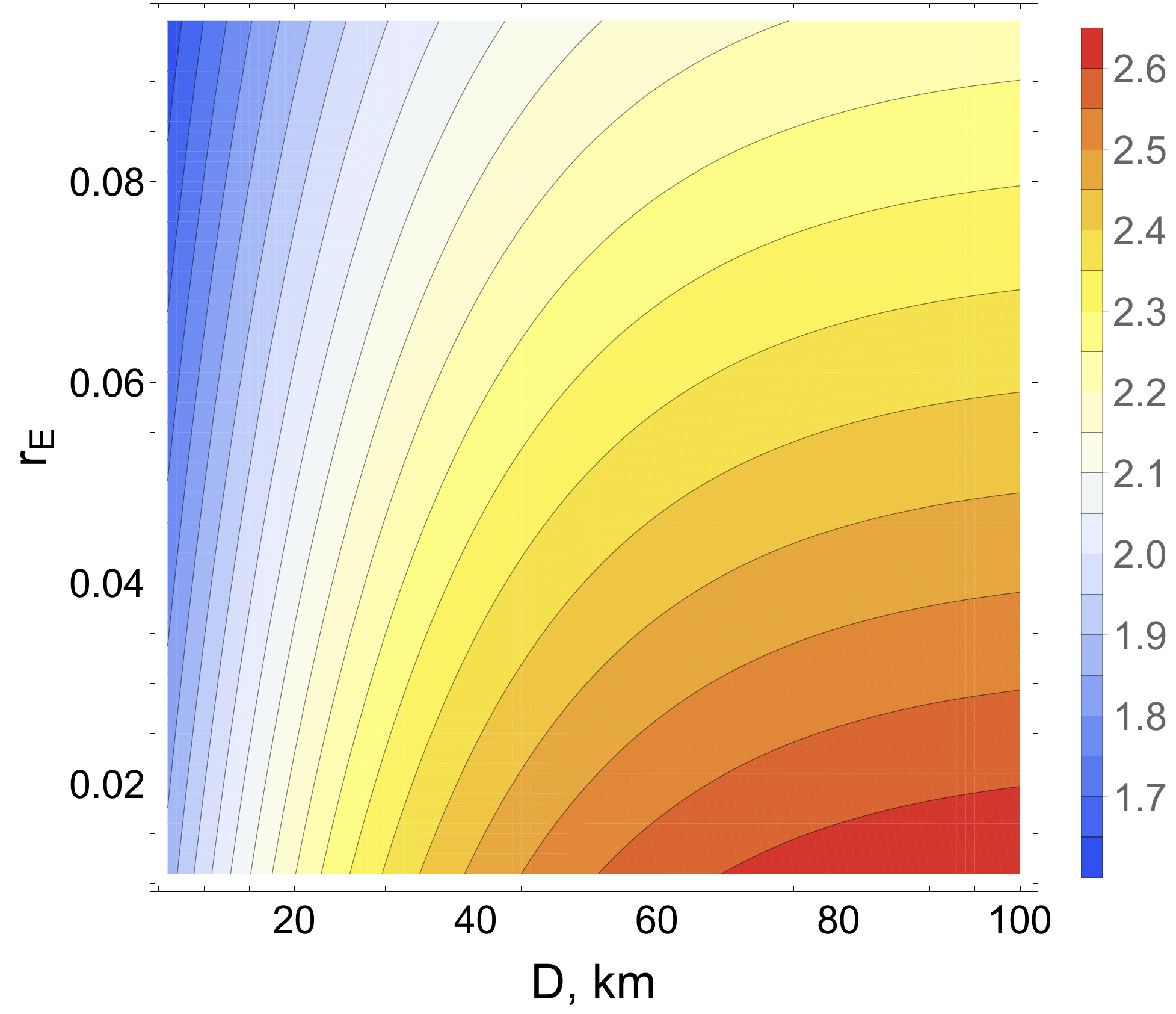} 
    \caption{The color plot for the ratio $R_{\mbox{\small{l.c.}}}/R_{\mbox{\small{decoy}}}$ between the rate for the BB84 protocol with line control $R=L'_f/L$ (\ref{BB84lc L}) and the key rate estimation for the original Decoy-State BB84 $R_{\mbox{\small{decoy}}}=\Tilde{L}_f/L$ (\ref{BB84 id Rkey}) as a function of $D$ and $r_E$ if the protocols parameters are preserved.}
    \label{bb84-ratio-no-intensity-optimisation}}
\end{minipage}
\end{figure}

Here, we compare the key rate estimation for the original BB84 with the Decoy-State method (\ref{BB84 id Rkey}), $R_{w.c.} = L_f/L$, and the key rate for the line controlled protocol (\ref{BB84lc L}), $R = L'_f/L$. 
The maximum of the upper bound\,(\ref{BB84 id Rkey}) is achieved with $|\gamma|^2=1$, i.e. when the pulses with one photon are the most possible. 
Let us first calculate the gain of the key generation rate if the line control is integrated only on the post-processing level, i.e., the physical parameters of pulses are not adapted to the level of losses.
It means that we are to put the $|\gamma|^2 = 1$ into the expression for the key generation rate (\ref{BB84lc L}).
Figure\,\ref{bb84-key-rate-no-intensity-optimisation} shows the dependence of the improved key rate on $D$ and $r_E$, Fig.\,\ref{bb84-ratio-no-intensity-optimisation} shows the ratio of the key rates for the improved and original protocols. 
Therefore, the line control technique gives us more than two times greater key generation rate even without any hardware modifications.

\begin{figure}[h]
    \centering
    \includegraphics[scale=0.4]{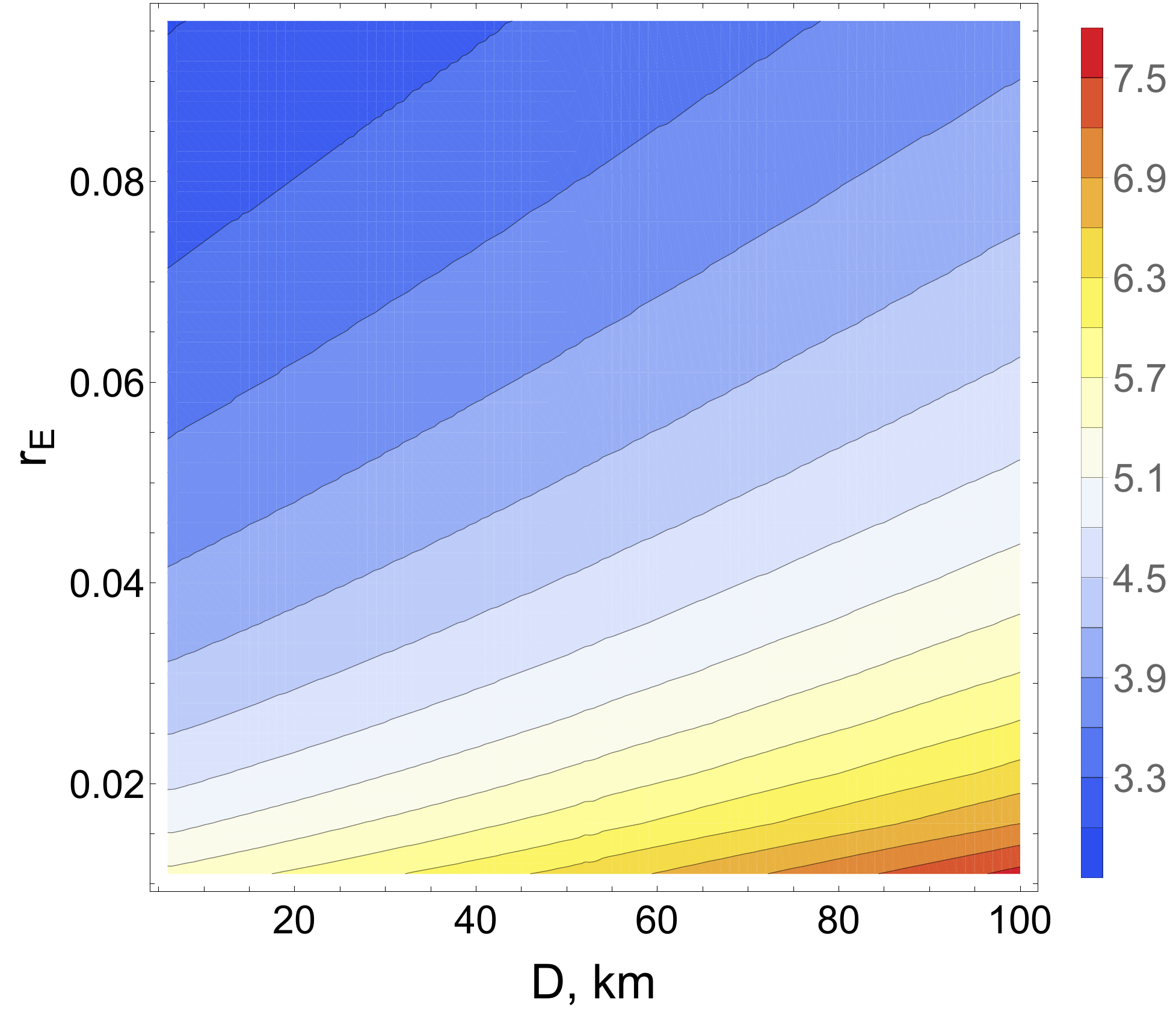}
    \caption{The color plot for the optimal intensity of the BB84 protocol with the line control as a function of $D$ and $r_E$; $\mu=1/50 \mbox{km}^{-1}$.}
    \label{cow-opt-int-fig}
\end{figure}

We can also numerically obtain the optimal intensity of the signal pulses maximizing the key rate (\ref{BB84lc L}) for different values of $D$ and $r_E$ -- the result is shown in Fig.\,\ref{cow-opt-int-fig}.
It can be seen that in the case of the line control, optimally, signals should comprise more than one photon.
For the 100 km-distance and leakage $\sim 10\%$, the optimal intensity is about 4 photons and increases up to 8 photons when the leakage is $\sim 1\%$. 
Thus, signal pulses better survive attenuation in the channel, and Eve cannot seize enough photons to successfully execute the PNS attack.

\begin{figure}[h]
\begin{minipage}[h]{0.47\linewidth}
    {\includegraphics
    [width=0.9\linewidth]{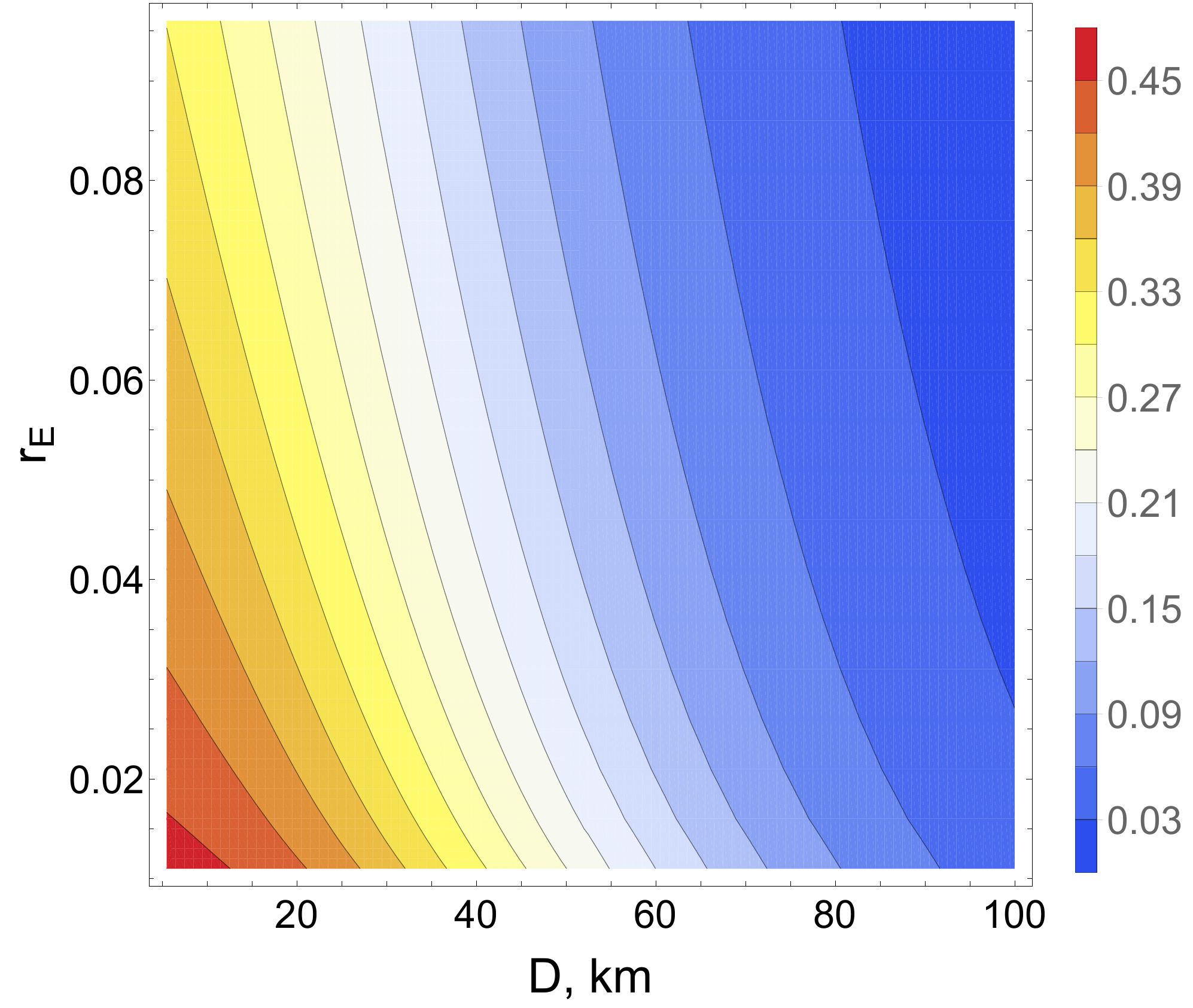}
    \caption{The color plot for the key rate (\ref{BB84lc L}) of the BB84 protocol with the line control as a function of $D$, the distance between Alice and Bob, and artificially crated leakage $r_E$, the signal loss caused by an eavesdropper; $\mu=1/50 \mbox{km}^{-1}$.}
    \label{BB84lc rkey}}
\end{minipage}
\hfill
\begin{minipage}[h]{0.47\linewidth}
    \center{\includegraphics
    [width=0.9\linewidth]{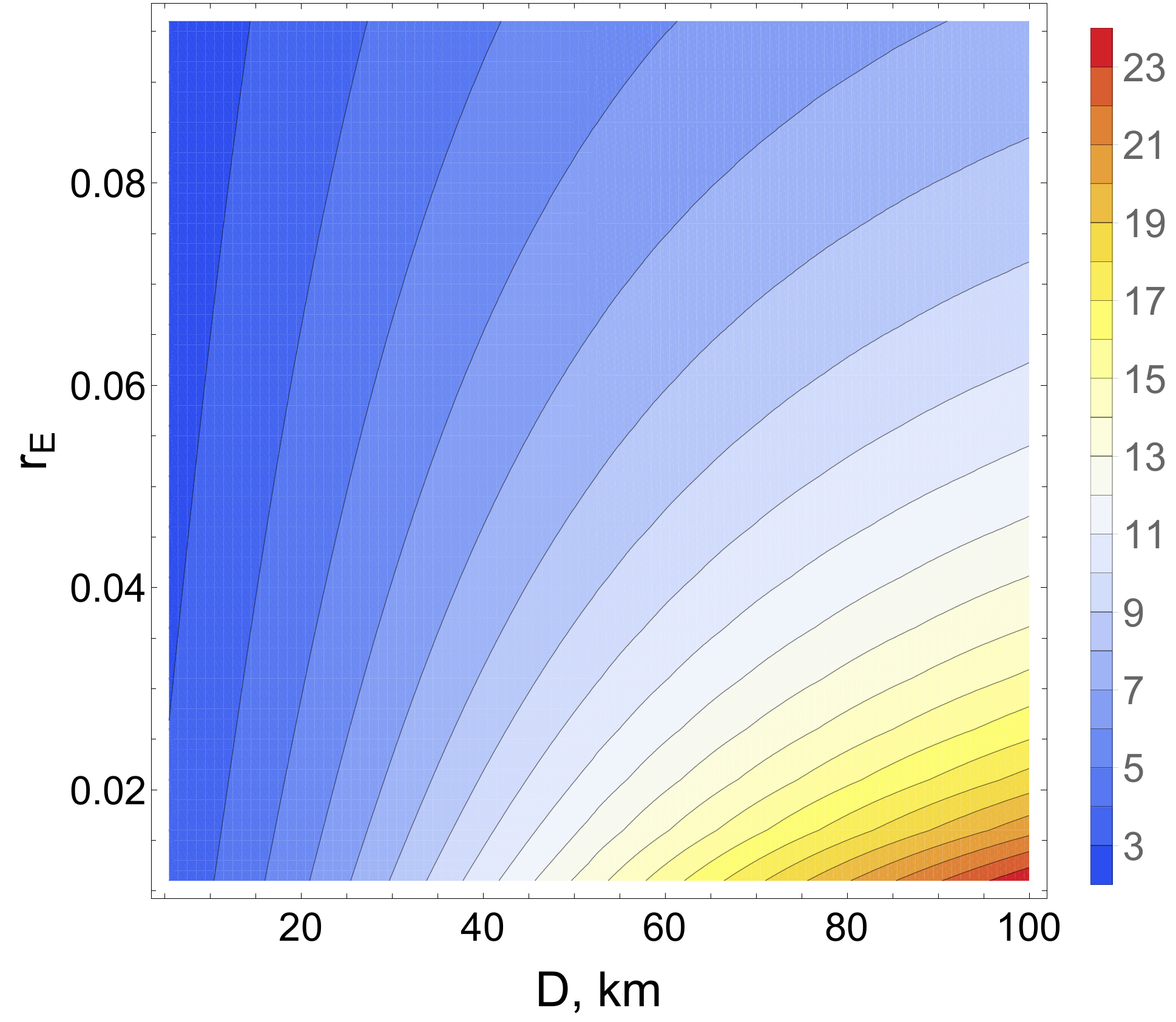} 
    \caption{The color plot for the ratio $R_{\mbox{\small{l.c.}}}/R_{\mbox{\small{decoy}}}$ between the rate for the BB84 protocol with line control $R=L'_f/L$ (\ref{BB84lc L}) and the key rate for the original Decoy-State BB84 $R_{\mbox{\small{decoy}}}=\Tilde{L}_f/L$ (\ref{BB84 id Rkey}) as a function of $D$ and $r_E$; $\mu = 1/50 \mbox{km}^{-1}.$
    }
    \label{BB84ratio}}
\end{minipage}
\end{figure}

Next, using the obtained optimal intensities we calculate the maximum key rate as a function of $r_E$ and the distance $D$ between Alice and Bob for the line controlled BB84, see Fig.\,\ref{BB84lc rkey}. 
The ratio between the key rate for the original Decoy-State BB84 and the key rate for the line controlled protocol as a function of $D$ and $r_E$ is represented in Fig.\,\ref{BB84ratio}. 

As shown, even the rough $10 \%$ leakage gives us several times greater key rate than in the original protocol.
If legitimate users can control Eve with $1 \%$ of the signal, they will benefit by getting more than 20 times larger key rate at the distance of 100\,km. 
Although this result is already impressive, the relative gain increases with for larger transmission distances.

\subsection{Improvement of the Coherent One-Way protocol}
Here, the Coherent One-Way (COW) protocol is discussed as an example of a protocol without phase randomization.
By analogy with the previous considerations, we first address the original version of the protocol and provide the upper estimate for the key generation rate modeling Eve's inception with the standard beam-splitter (BS) attack in which she gets all dissipative losses.
Secondly, we calculate the key rate for the modified protocol with the line control.
Owing to the efficient line control, the losses associated with the eavesdropper can be separated allowing thus, for the less destructive reduction of the key length during the privacy amplification stage.

Note that in both versions of the protocol, the intensities of the signal pulses are adjusted to maximize the key rate.

\subsubsection{Original COW}

To begin with, let us consider the standard COW QKD protocol, in which Alice utilizes an attenuated laser and prepares a coherent state with intensity $|\gamma|^2$ to encode a random bit string into two-pulse sequences composed of non-empty and empty pulses:
\begin{gather}
    \mbox{bit }``0" \quad\longleftrightarrow\quad \ket{0}\ket{\gamma},
    \qquad
    \mbox{bit }``1" \quad\longleftrightarrow\quad \ket{\gamma}\ket{0}.
\end{gather}

\begin{figure}[h]
    \centering
    \includegraphics[scale=0.8]{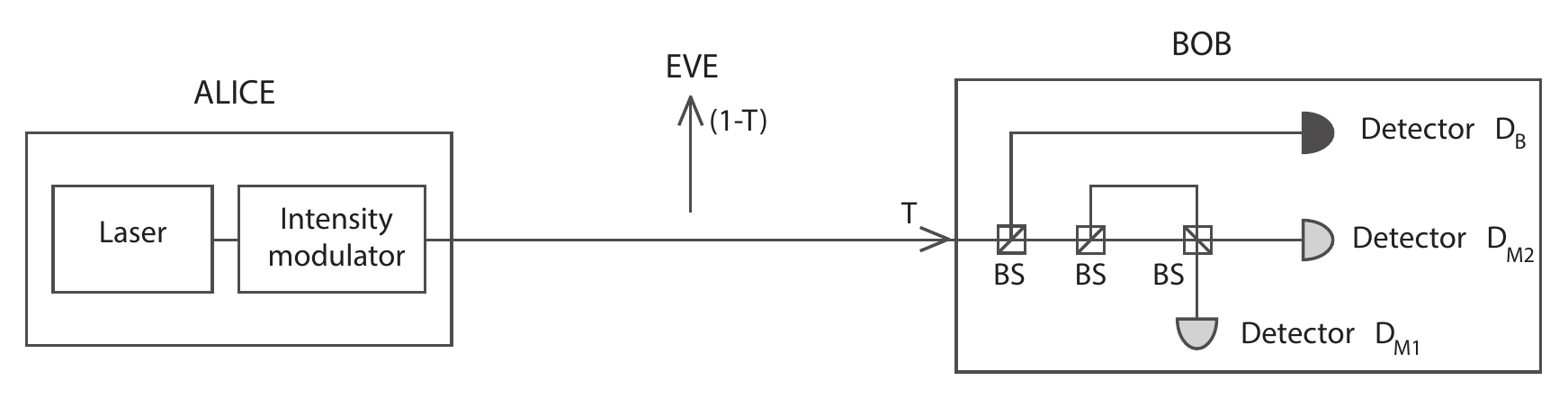}
    \caption{The scheme of the COW QKD protocol. Alice prepares coherent states using the laser and adjusts the signal's intensity by the intensity modulator. Bob utilizes a single-photon detector $D_B$ to monitor the signal's arrival time. Single-photon detectors $D_{M1}$ and $D_{M2}$ check whether the arriving signal is a decoy state. Eve obtains all dissipative losses $1-T$.}
    \label{cow_scheme}
\end{figure}

A small fraction $f \ll 1$ of all two-pulse sequences are the decoy states $\ket{\gamma}\ket{\gamma}$. 
The long interferometer's arm (see Fig. \ref{cow_scheme}) has the length assuring that two non-empty adjacent pulses interfere at the last beam-splitter. 
The detector $\mbox{D}_{M2}$ does not react to the decoy states and the sequences with two subsequent $\ket{\gamma}$ states like 0-1 sequences.
Legitimate users estimate the visibility of the interference in the detectors $\mbox{D}_{M1}$ and $\mbox{D}_{M2}$ and use it to estimate the eavesdropper's information. 
The main detector $D_B$ monitors the time of arrival of the pulse. 
Sometimes $D_B$ will not react to non-empty pulses, because of the Poisson photon-number statistics governing the coherent pulses. 
Bob interprets these measurement results as inconclusive. 
After transmitting all pulses, Alice announces which states were decoy and which were information-carrying. 
The post-selection procedure involves identifying decoy pulses and discarding the positions at which Bob obtained inconclusive results. 
A potential eavesdropper may introduce some additional errors. 
Consequently, Bob will obtain more inconclusive results than he would expect as a result of the losses in the channel. 

To give an upper estimate of the key rate in the original protocol, it will be sufficient to consider any of the possible eavesdropping attacks. 
For instance, one can analyze the attack in which Eve obtains the lost part of the signal. 
If $D$ is the distance between Alice and Bob, the transmittance of the whole optical line is determined by $T = 10^{-\mu D}$ ($\mu=1/50 \mbox{km}^{-1}$). 
The maximum Eve's information about the sent bit can be estimated as the Holevo bound \cite{Holevo}. 
For an ensemble $\left\lbrace p_i, \rho_i \right\rbrace_{i}$ it is defined as $\chi = S\left( \sum_i p_i \rho_i \right) - \sum_i p_i S \left( \rho_i \right)$, where $S(\rho) = - \mbox{Tr}\left( \rho \log \rho \right)$ is the von Neumann entropy.
For the equiprobable pure states the Holevo value has the from
\begin{gather}
    \max I(A,E)=\chi\left(\ket{\sqrt{1-T}\cdot\gamma}\otimes\ket{0},\ket{0}\otimes\ket{\sqrt{1-T}\cdot\gamma}\right)=h_2\left(\frac{1-|\braket{\sqrt{1-T} \cdot\gamma|0}|^2}{2}\right)\,,
    \label{COW Eve}
\end{gather}
where the binary entropy $h_2$ was defined after Eq.\,(\ref{DecoyKey}).
The probability of the conclusive measurement result on the Bob's side is
\begin{gather}
    p\left( \checkmark \right)=1-\exp\left(-10^{-\mu D} \cdot |\gamma|^2\right).
    \label{COW conclusive}
\end{gather}

To eliminate  the  eavesdropper  information, Alice  and  Bob  perform key distillation procedure, for  instance, agree (through  a public  authenticated  channel  or,  alternatively,  before  the  protocol  is  executed)  on  a  random  function $g:\{ 0,1 \}^{p(\checkmark)L} \rightarrow \{ 0,1\}^{L_f}$ shrinking a shared bit string of the length $p(\checkmark) \cdot L$, which they have after the post-selection procedure, to the reduced size 
\begin{gather}
    L_f
    =
    p\left( \checkmark \right) L \cdot \left(1-\max I \left( A,E \right) \right)
    = p\left( \checkmark \right) L \cdot\left(1-h_2\left(\frac{1-\exp\left(-(1-T) \cdot |\gamma|^2 \right) }{2}\right)\right).
    \label{id}
\end{gather}

\subsubsection{The COW improved by the line control}

Let us consider artificial losses caused by an eavesdropper in addition to the channel losses, in the case where the physical control over the transmission line is realized. 
The probability for the measurement result to be conclusive is
\begin{gather}
    p' \left( \checkmark \right)=1 - \exp\left(-10^{-\mu D} \cdot (1-r_E) \cdot |\gamma|^2\right),
    \label{COW lc p}
\end{gather}
where $r_E$ is the portion of signal intensity that an eavesdropper gets. 
To estimate the maximum information that Eve can gain, we again calculate the Holevo bound
\begin{gather}
    \max I' (A,E)=\chi\left( \ket{\sqrt{r_E} \gamma}\otimes\ket{0},\ket{0}\otimes\ket{ \sqrt{r_E} \gamma} \right).
    \label{COW lc I}
\end{gather}
After the post-selection procedure and privacy amplification, a shared bit string has the size
\begin{gather}
    L'_f
    =
    p' \left( \checkmark \right)L\cdot\left(1-\max I' \left( A,E \right) \right)
    =
    p' \left( \checkmark \right)L\cdot\left(1-h_2\left(\frac{1-\exp\left( -r_E \cdot|\gamma|^2 \right)}{2}\right)\right).
    \label{c}
\end{gather}

\begin{figure}[h]
\begin{minipage}[h]{0.47\linewidth}
    \center{\includegraphics
    [width=0.9\linewidth]{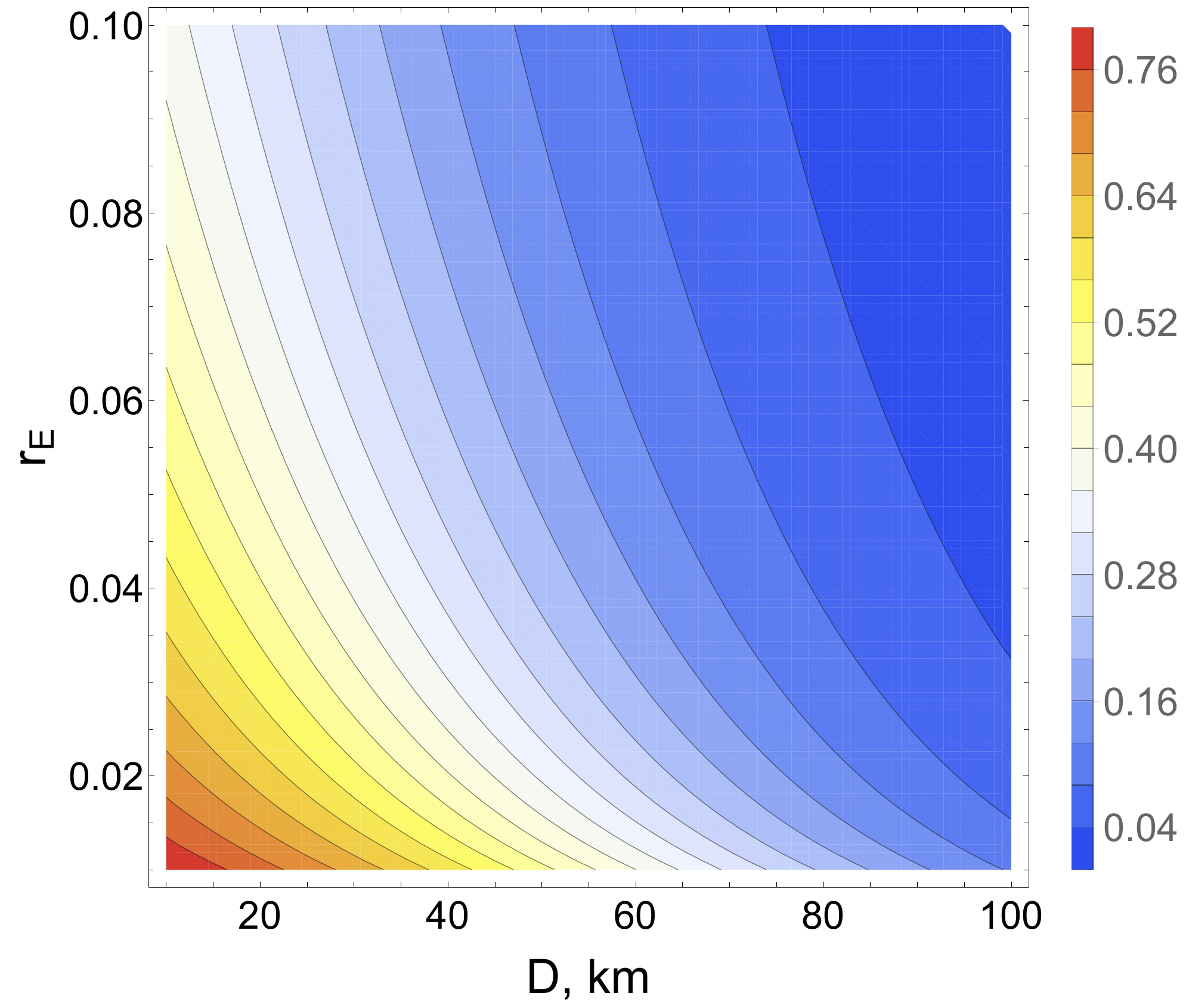}
    \caption{The color plot for the key rate of COW with line control as a function of $D$, the distance between Alice and Bob, and $r_E$, the signal loss caused by an eavesdropper, with $\mu=1/50 \mbox{km}^{-1}$.}
    \label{keyrate_COW_plot}}
\end{minipage}
\hfill
\begin{minipage}[h]{0.47\linewidth}
    \center{\includegraphics
    [width=0.9\linewidth]{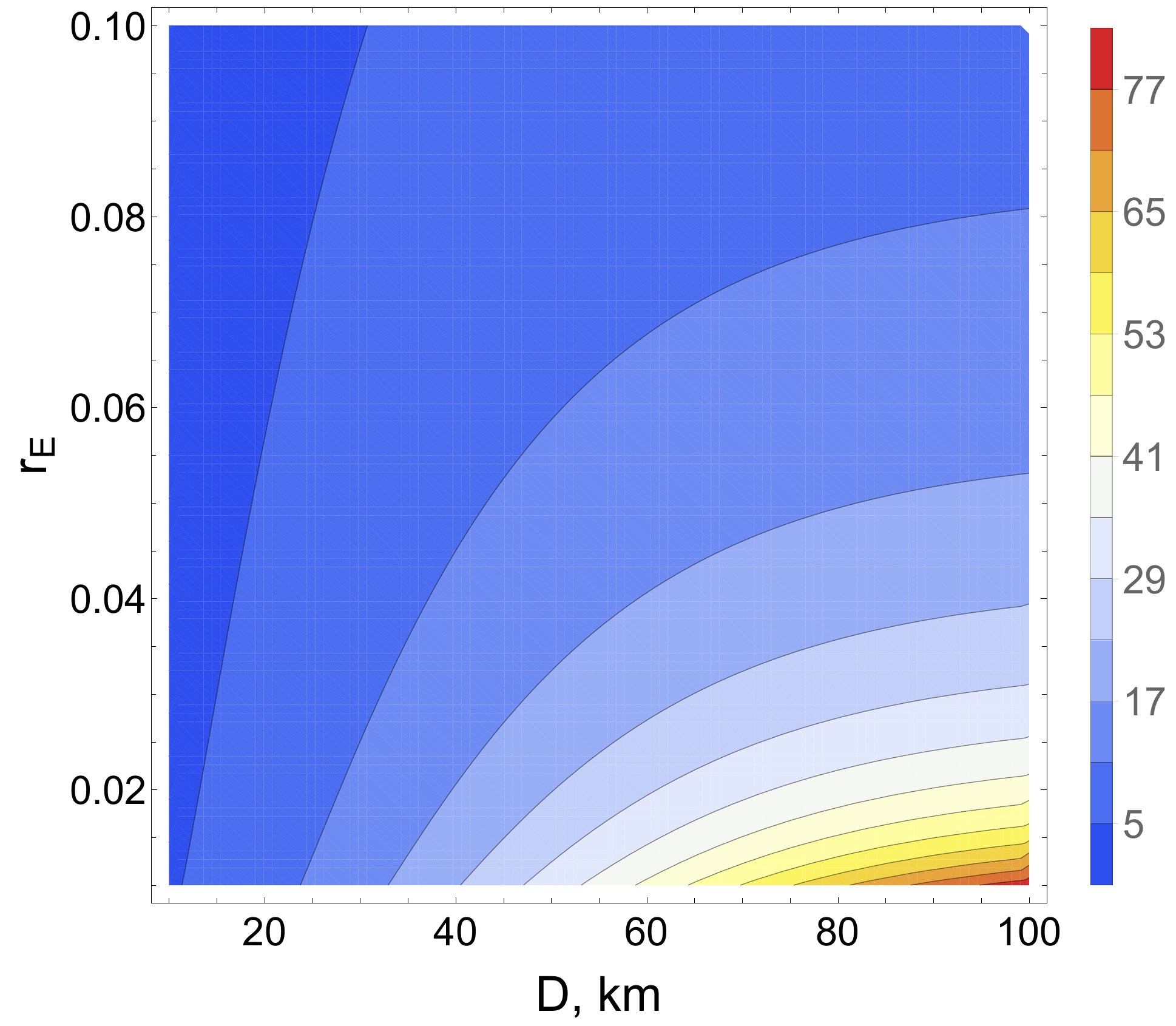} 
    \caption{The color plot for the ratio $R/R_{\mbox{w.c.}}$ between the key rate for the COW protocol with the line control $R=L'_f/L$ and the key rate for original COW $R_{\mbox{w.c.}}=L_f/L$; $\mu=1/50 \mbox{km}^{-1}$.}
    \label{fraction_plot}}
\end{minipage}
\end{figure}

Next, we compare the key rate estimation for the original COW (\ref{id}), $R_{\mbox{w.c.}}=L_f/L$ and the key rate of the line controlled protocol (\ref{c}), $R=L'_f/L$.
Each protocol has its optimal intensity $|\gamma|^2$ which provides the maximum key rate for the particular attenuation magnitude $10^{- \mu D}$.  
For different values of $r_E$, we have obtained optimal intensities by numerically finding the maximum of (\ref{id}) and (\ref{c}). 
By substituting the intensity optimal for $10^{-\mu D}$ in formulas (\ref{id}) and (\ref{c}) we calculate the maximum key rate as a function of $r_E$ and $D$ for the original COW and the line controlled protocol.
Fig. \ref{keyrate_COW_plot} represent the key rate for the line controlled protocol, Fig. \ref{fraction_plot} represents the ratio between the key rates.

We see that the modified protocol always produces substantially higher key rates than the original COW even in the pessimistic case where Eve gets $10\% $ of the signal.
Note that, as we shown previously, the loss detection precision enables to notice much subtle leakages: if the line control shows that Eve seizes $1\%$ of the signal, modified protocol gives $\sim70$ times larger key rate than the original one.

In the case of the COW protocol, line control implementation means that legitimate users do not have to use the decoy states and analyze a large number of bits corresponding to decoy pulses. 
Thus, the protocol's framework can be significantly simplified by removing the interference part of the scheme with detectors $D_{M1}$ and $D_{M2}$. 
These modifications may reduce the cost of commercial offers and make the QKD implementations more market-available.

\subsection{Differential phase shift QKD protocol}
\begin{figure}[h]
    \centering
    \includegraphics[scale=0.8]{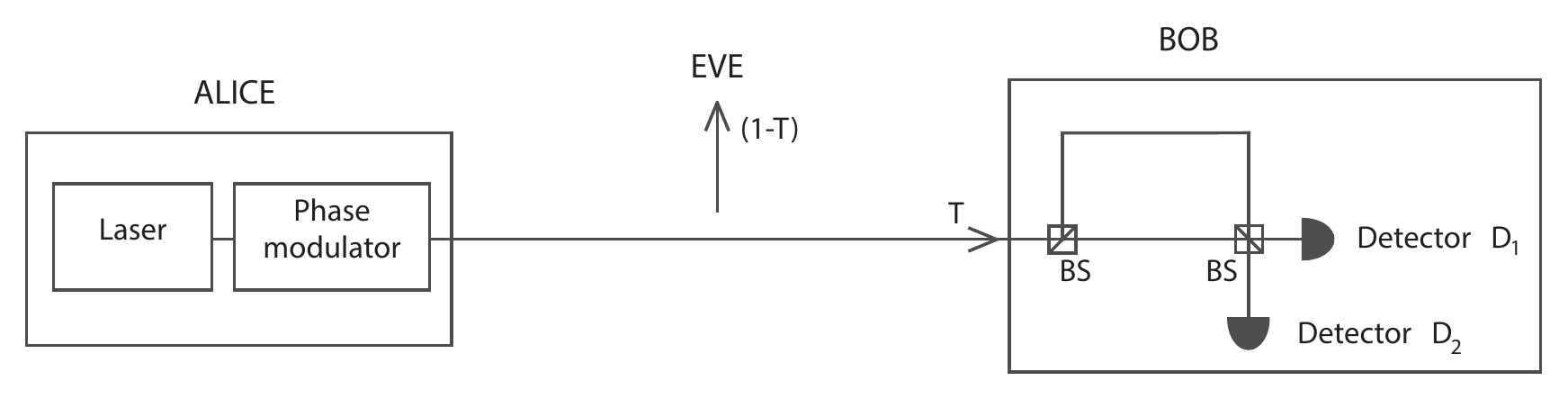}
    \caption{The scheme of the DPS QKD protocol. 
    Alice prepares coherent states by a laser and shifts the signal's phase by a phase modulator. 
    Bob's interferometer consists of two beam-splitters BS and two single-photon detectors $D_{M1}$ and $D_{M2}$ to checks whether the phase difference between two adjacent pulses is $0$ or $\pi$. 
    Eve obtains all dissipative losses of a fraction $1-T$. 
    See text for details.}
    \label{DPS_sheme}
\end{figure}

Let us briefly demonstrate the work of our method in the case of the Differential Phase Shift (DPS) protocol.
In the DPS QKD protocol, see Fig. \ref{DPS_sheme}, Alice uses weak coherent pulses that are randomly phase-modulated by $\left\lbrace 0, \pi \right\rbrace$ for each time bin.
Each incoming pulse splits between the two paths. 
Then Bob recombines them by 50:50 beam-splitters, where the long interferometer’s arm has such a length that two adjacent pulses interfere at the last beam-splitter.

This setup allows for the partial wave functions of two sequential pulses to interfere with each other. 
Detector 1 clicks for 0 phase difference between the two consecutive pulses and detector 2 clicks for $\pi$ phase difference. 
After receiving pulses, Bob announces to Alice the times of pulses' detection. 
Given the information about the time and modulation, Alice knows which of the Bob's detectors should have clicked. 
By denoting the detector's 1 (2) click as ``0'' (``1'') Alice and Bob obtain a raw key.
\begin{gather}
    \mbox{bit ``0"} \quad\longleftrightarrow\quad \ket{\gamma}\ket{\gamma} \mbox{ or } \ket{-\gamma}\ket{-\gamma},
    \qquad
    \mbox{bit``1"} \quad\longleftrightarrow\quad \ket{\gamma}\ket{-\gamma} \mbox{ or } \ket{-\gamma}\ket{\gamma}.
\end{gather}
Considering the attack in which Eve obtains the lost part of the signal, one may notice that formula (\ref{COW Eve}) for the DPS case transforms into
\begin{gather}
    \max I(A,E)=\chi\left(\ket{\sqrt{1-T}\cdot\gamma}\otimes\ket{-\sqrt{1-T}\cdot\gamma},\ket{\sqrt{1-T}\cdot\gamma}\otimes\ket{\sqrt{1-T}\cdot\gamma}\right)=\\
    \nonumber
    =h_2\left(\frac{1-|\braket{\sqrt{1-T} \cdot\gamma|-\sqrt{1-T}\cdot\gamma}|^2}{2}\right).
\end{gather}
Meanwhile the structure of the formula (\ref{COW conclusive}) for $p(\checkmark)$ is preserved. 
Finally, Eq.\,(\ref{id}) evolves into
\begin{gather}
    L_f = p(\checkmark)L \cdot \left( 1-h_2 \left( \frac{1-\exp\left( -4 \cdot (1-T)\cdot |\gamma|^2 \right)}{2} \right) \right).
    \label{DPS_kgr}
\end{gather}
Considering the line control procedure for the DPS protocol, the probability of a conclusive measurement result is precisely the same as for the COW (\ref{COW lc p}), so that Eqs.\,(\ref{COW lc I}) and (\ref{c}) become
\begin{gather}
    \max I'(A,E) = \chi \left( \ket{\sqrt{r_E}\cdot\gamma}\otimes\ket{-\sqrt{r_E}\cdot\gamma},\ket{\sqrt{r_E}\cdot\gamma}\otimes\ket{\sqrt{r_E}\cdot\gamma} \right),\\
    L'_f
    =
    p' \left( \checkmark \right)L\cdot\left(1-\max I' \left( A,E \right) \right)
    =
    p' \left( \checkmark \right)L\cdot\left(1-h_2\left(\frac{1-\exp\left( -4 \cdot r_E \cdot|\gamma|^2 \right)}{2}\right)\right).
    \label{DPS_lc_kgr}
\end{gather}

Since the equations for the DPS and COW protocols are nearly identical, similar analysis can be performed to determine the optimal signal intensities, key rates, and the ratio between the key rate for the line controlled protocol (\ref{DPS_lc_kgr}) and the key rate estimation for the original DPS's (\ref{DPS_kgr}). 
Although the particular values will be different, the characteristic behavior of the key rate gain will be the same.

\section{Discussion}
In summary, we have proposed a method that can enhance any prepare \& measure recipe for the Quantum Key Distribution protocol (i.e. to significantly increase a key rate and an achievable distance between the legitimate users). 
The method involves the physical control over the transmission line directed at detecting any intrusion of an eavesdropper. 
The physical control comprises measurements carried out on the optical reflectometer, based on the registration of the back-scattered optical radiation, and the cross-checking parameters in the test pulses measured by Bob and prepared by Alice.

We have classified all QKD protocols with optical realisations into two groups and demonstrated for each group the benefits of using the proposed line control technique. 
The first group consists of the protocols which were originally designed for exploiting single-photon states.
Namely, we have addressed the BB84 protocol with the Decoy-State method and demonstrated the increase in the key generation rate by a factor of 20, at the distance of 100\,km. 
Secondly, we have considered protocols that utilize pure coherent states as signal pulses. 
Using the COW as an exemplary protocol, we showed that one may get about 70 times larger key rate over the distance of 100\,km. 
We have demonstrated that the line control implementation allows to simplify physical realizations of some protocols, which, in turn, reduces the demand on the equipment. 
Moreover, even when implemented on the post-selection level, without any changes in signal preparations, our method significantly boosts the key distribution.

\printbibliography
\end{document}